\input epsf 
\documentstyle[preprint,epsfig,eqsecnum,aps]{revtex}

\begin{document}

\count255=\time\divide\count255 by 60 \xdef\hourmin{\number\count255}
  \multiply\count255 by-60\advance\count255 by\time
 \xdef\hourmin{\hourmin:\ifnum\count255<10 0\fi\the\count255}

%\draft
%\preprint{\vbox{\hbox{WM-00-101}\hbox{JLAB-THY-00-04}}}

\title{Generations of Higgs Bosons in Supersymmetric Models}

\author{Alfredo Aranda\footnote{fefo@physics.wm.edu} and Marc Sher\footnote
{sher@physics.wm.edu}}

\vskip 0.1in

\address{Nuclear and Particle Theory Group, Department of
Physics, College of William and Mary, Williamsburg, VA 23187-8795}
\vskip .1in
\date{May, 2000}
\vskip .1in

\maketitle
\tightenlines
\thispagestyle{empty}

\begin{abstract}
We examine extensions of the MSSM with more than one generation of Higgs bosons. If
one assumes that a symmetry eliminates the tree-level FCNC, then the extra scalar
bosons do not acquire VEVs, do not couple to fermions and do not mix with the ordinary
Higgs bosons; the lightest is absolutely stable. The  
two lightest neutral scalars, $\phi_S$, and $\phi_P$, are degenerate in mass, and
the mass difference between those and the  lightest charged scalar, $\phi_+$, is
calculated. For most of the parameter space,  the charged scalar is between a hundred MeV and a few GeV heavier
than the neutral scalars.  The $\phi_+$ will decay
at the vertex; the signature for this decay will be like that of a chargino with a
nearly degenerate undetected neutralino.
Next, the possibility that the symmetry that eliminates FCNC is a flavor symmetry is
discussed. In an example, the U(2) model, tree-level FCNC processes can be calculated
in terms of quark masses. The strongest constraint on this model is from $D-\bar{D}$
mixing, which should be within an order of magnitude of the current bound.
\end{abstract}

\newpage
\setcounter{page}{1}

\section{Introduction}\label{section1}
One of the great mysteries of the standard model is the number of
fermion generations. Nothing in the structure of the standard model gives any 
clue as to what this number should be, and the same is true of the most popular
extensions of the standard model $-$ grand unified theories and the MSSM. 
The only information we have about the number of fermion generations is
phenomenological. It is 
interesting that this situation is the same for the Higgs bosons, i.e., there
is no clue as to the number of generations of Higgs bosons (
in supersymmetric models each generation
consisting of two doublets of opposite hypercharge). One might \emph{a priori}
expect the number of Higgs generations to be equal to the number of fermion
generations. This is because in some supersymmetric grand unified theories,
such as $E_6$, the Higgs bosons and fermions belong to the same representations.
Just as in the fermionic sector, the number of Higgs generations can, at present,
only be determined phenomenologically.

Of course, the number of Higgs generations detected experimentally 
at present is zero.
However, a strong clue comes from the
absence of tree-level flavor-changing neutral currents (FCNC). In the standard
model with a single Higgs doublet, the Yukawa coupling matrices are
proportional to the fermionic mass matrices. Diagonalizing the latter thus 
automatically diagonalizes the former, eliminating FCNC. If one or more Higgs
doublets are added to the standard model, diagonalizing the mass matrices does not
in general diagonalize the Yukawa coupling matrices. In minimal supersymmetric
extensions of the standard model, there are two Higgs doublets of 
opposite hypercharge, one couples to the $Q=2/3$ quarks while the other
couples to the $Q=1/3$ quarks, thereby eliminating FCNC. Again, with an
extra pair of doublets added to these minimal models, FCNC naturally emerge.

Since we are interested in the possibility of having additional 
generations of Higgs bosons, it is important to solve
the problem of FCNC~\cite{higgsdoublets}. 
There are two approaches\footnote{Although generally discussed
in the context of two-generation models, these approaches apply more generally.}.
The first~\cite{glawein} eliminates them completely by imposing a discrete
symmetry. The second approach~\cite{chengsher} does not eliminate FCNC, but
makes them sufficiently small by assuming that the flavor-changing neutral 
couplings of quarks $q_i$ and $q_j$ are proportional to the geometric mean of
the Yukawa couplings of the same quarks. In this case, the most dangerous
FCNC, which involve the down and strange quarks, are suppressed by the
small down and strange Yukawa couplings\footnote{Another possible solution
to the FCNC may be that the additional Higgses are too heavy, and thus the
effective low energy
theory is the usual MSSM.}.

In Ref.~\cite{griestsher}, the first approach (applying a discrete symmetry
to remove FCNC) was applied to the possibility of adding extra generations 
of Higgs bosons to the MSSM. They considered three generations of Higgs bosons:
six doublets, three of each hypercharge. It was shown that if one assumes
that some symmetry (discrete, continuous, global or local) suppresses FCNC,
then the extra generations, which they called ``pseudoHiggs bosons'' do not 
acquire vacuum expectation values, do not mix with the ``standard'' doublets 
and do not couple to fermions. The lightest of these extra scalars is stable, 
and much of Ref.~\cite{griestsher} was devoted to the possibility that this scalar
could constitute the dark matter.

In this article, we analyze the MSSM with extra generations of Higgs doublets.
Initially, we assume that there is a symmetry which eliminates tree-level
FCNC 
(at this point, we do not worry about
the precise nature of the symmetry). In Section~\ref{section2}, we present
the model and show that the lightest neutral scalars are degenerate in mass, and that the
lightest charged scalar is only a few GeV heavier. It is shown that standard searches for
heavy leptons(charginos) with neutrinos(neutralinos) a few GeV lighter will  be sensitive to these particles,  In
Section~\ref{section3} we relax the assumption that tree-level FCNC are eliminated, and
consider a simple flavor symmetry, based on U(2), and show how the resulting tree-level
FCNC can be calculated in terms of fermion masses. 
\section{The Model}\label{section2}
Let us first summarize the model discussed in Ref.~\cite{griestsher}. We consider
the supersymmetric standard model with three generations of Higgs doublets. The
most general superpotential is given by
\begin{eqnarray}\label{W}
W=\mu_{ij} H_i \bar{H}_j + f_{ijk}Q_iU_jH_k +g_{ijk}Q_iD_j\bar{H}_k +
h_{ijk}L_iE_j\bar{H}_k \, ,
\end{eqnarray}
where the lowercase Latin indices run over 1,2,3. It is assumed that some
symmetry eliminates the tree-level FCNC in the quark and lepton sectors. Given
this assumption, a basis can be chosen in which only one generation, 
conventionally chosen to be $H_3$ and $\bar{H}_3$, couples to quarks and
leptons. This means that $H_3$ and $\bar{H}_3$ must have different quantum
numbers under that symmetry than the other $H_{1,2}$ and $\bar{H}_{1,2}$, in
order that only the former couple to fermions.

The most general soft SUSY-breaking terms involving only the Higgs fields are
\begin{eqnarray}\label{soft}
W_{soft} \supset m_{H_{ij}}^2 H_{i}^{\dagger} H_{j} + 
 m_{\bar{H}_{ij}}^2 \bar{H}_{i}^{\dagger} \bar{H}_{j} -
B \mu_{ij} \left( H_{i} \bar{H}_{j} + h.c. \right).
\end{eqnarray}
Since $H_3$ and $\bar{H}_3$ have different quantum numbers 
than the other $H_{1,2}$ and $\bar{H}_{1,2}$ under the symmetry,
then the quadratic terms involving only one of them vanish, i.e. 
$m_{H_{i3}}^2 = m_{\bar{H}_{i3}}^2 = m_{H_{3i}}^2 = m_{\bar{H}_{3i}}^2 =
\mu_{i3} = \mu_{3i} = 0$ for $i = 1, 2$. Thus there are no quadratic terms
mixing the third generation of Higgs fields with the other two. In addition,
Ref.~\cite{griestsher} shows that equality of scalar masses at the unification
scale automatically implies that the first and second generation fields do
not get vacuum expectation values. Note that we have four new scalar fields,
four new pseudoscalar and four new pairs of charged scalars.
In the following, we specialize to the case in which there is only a single
generation of extra Higgs fields, denoted by $H_X$ and $\bar{H}_X$. This is 
primarily for simplicity$-$including the additional generations does not
affect our results. In fact, if the two additional generations have different 
quantum numbers under the symmetry that eliminates FCNC, then the generations
will decouple and this specialization is completely general. Even if the
coupling between the two generations exists, however, the results we
present below are completely unaffected$-$the results would then simply
apply to the lightest of the scalar fields.

So, assuming only one extra generation of Higgs fields, the scalar potential
can be written as
\begin{eqnarray}\label{V}
\nonumber
V & = & m_X^2 |H_X|^2 + \bar{m}_X^2 |\bar{H}_X|^2 + m_1^2 |H|^2 + 
m_2^2 |\bar{H}|^2 + \left( \mu_X H_X \bar{H}_X +\mu H \bar{H} + h.c. \right)\\
\nonumber
  & + & \frac{g^2}{8} \sum_a \left| H_X^{\dagger} \tau_a H_X + \bar{H}_X^{\dagger}
\tau_a \bar{H}_X + H^{\dagger} \tau_a H + \bar{H}^{\dagger} 
\tau_a \bar{H} \right|^2 \\
  & + & \frac{g^{'2}}{8} \left| |H|^2 + |H_X|^2 - |\bar{H}|^2 - 
|\bar{H}_X|^2 \right|^2 \, ,
\end{eqnarray}
where $H$ and $\bar{H}$ are the standard MSSM doublets, $\tau_a$ are the Pauli
matrices, and where $\mu$ and $\mu_X$ are arbitrary parameters of dimension
$(\rm{GeV})^2$. The full Lagrangian has a symmetry under which the $H_X$ fields
change sign (it is actually a global U(1) symmetry), and thus the lightest of
these fields is stable.

The mass matrices of the scalars can be calculated. The mass matrix for the 
additional neutral scalar fields is given by
\begin{eqnarray}\label{massscalar}
M_S^2 = \left( \begin{array}{cc} m_X^2 -\frac{1}{2} M_Z^2 \cos 2 \beta & -\mu_X \\
-\mu_X & \bar{m}_X^2 +\frac{1}{2} M_Z^2 \cos 2 \beta \end{array} \right) \, ,
\end{eqnarray}
and the matrix for the additional neutral pseudoscalars is given by
\begin{eqnarray}\label{masspseudo}
M_P^2 = \left( \begin{array}{cc} m_X^2 -\frac{1}{2} M_Z^2 \cos 2 \beta & \mu_X \\
\mu_X & \bar{m}_X^2 +\frac{1}{2} M_Z^2 \cos 2 \beta \end{array} \right) \, .
\end{eqnarray}
The mass matrix for the charged scalars is
\begin{eqnarray}\label{masscharged}
M_{+}^2 = \left( \begin{array}{cc} m_X^2 +\frac{1}{2} M_Z^2 \cos 2 \bar{\beta} & 
-\mu_X \\ -\mu_X & \bar{m}_X^2 -\frac{1}{2} M_Z^2 \cos 2 \bar{\beta} \end{array} 
\right) \, ,
\end{eqnarray}
where $\cos 2 \bar{\beta} \equiv \cos 2\beta \cos 2\theta_W$. Note that the mass
matrices for the scalar and pseudoscalar are \emph{identical} except for the
sign of the even-odd elements (this is true even in the case of many additional
generations all coupled together). As a result, the secular equation is identical
for both mass matrices, and so the eigenvalues are the same.  The charged scalar mass
matrix is identical to the neutral scalar mass matrix with $M_Z^2 \rightarrow M_Z^2 \cos
2\theta_W$. This results in a \emph{larger} mass for the lightest charged scalar, but only
slightly larger$-$we will see that a few GeV is a typical size.

None of this is new. It was discussed in much more detail in Ref.~\cite{griestsher}.
However,  there was an error in that paper (that also appeared in an earlier version of
this paper).  They argued that the $S-P$ mass degeneracy would be split by radiative 
corrections
involving a loop with a $W$ boson, noting a difference between the magnitude of the
couplings of the $\phi_S$ to the $W$ and charged scalar and the magnitude of the couplings
of the $\phi_P$.  However, this difference was due to a sign error in a Feynman rule; the
couplings are of the same magnitude.  Thus the degeneracy is not split by radiative
corrections.  One can see this by noting that the $U(1)$ global symmetry which ensures the
stability of the lightest scalar has the $H_X$ with charge +1, the $\bar{H}_X$ with charge
-1 and the rest with charge zero.  This automatically gives the mass degeneracy, and this
global symmetry is unbroken.  In Ref.~\cite{griestsher}, there was no discussion of the mass
difference between the charged and neutral scalars, nor the decay modes.  We now look at
this mass difference, and discuss the signatures.

The mass matrices Eq.~(\ref{massscalar}),(\ref{masspseudo}), and (\ref{masscharged})
have four unknown parameters: $m_X^2$, $\bar{m}_X^2$, $\mu_X$, and $\tan\beta$.
However, as noted above, the beta functions for $m_X^2$ and $\bar{m}_X^2$ are
identical, and equality of the masses at a high scale implies that they are equal
at all scales. We thus set them equal to each other (our results are not significantly affected by relaxing this assumption).  We also see that the results are very insensitive to $\tan\beta$. Thus, there are effectively two parameters. These two
parameters give the masses of the  neutral scalars, 
and the two charged scalar pairs, as well as all of the mixing angles. The 
coupling to the vector bosons are thus determined in terms of these parameters, and
are given in Ref.~\cite{griestsher} (but note that, as discussed in the previous
paragraph, the Feynman rule involving $\phi_P,\phi_+$ and the $W$ has a sign error in
front of the $\theta_+$ term).

We now turn to the mass difference between the lightest charged Higgs and $\phi_S$.
  At tree level, the splitting can be
easily found from the mass matrices and is given by
\begin{equation}
M^2_+-M^2_S= {1\over 2}\left(
\sqrt{M_Z^4\cos^22\beta+4\mu_X^2}-\sqrt{M_Z^4\cos^22\bar{\beta}+4\mu^2_X}\right)
\end{equation}  
Numerically, for $\tan\beta=2$, the mass splitting  is ${150\
{\rm GeV}\over M_S}$ times $(2.9,0.7,0.2)$ GeV for $\mu_X^{1/2}=(40,100,200)$ GeV.
For $\tan\beta=10$, the splitting for the same $\mu_X$ values is ${150\ {\rm GeV}\over
M_S}$ times $(5.5,1.9,0.6)$ GeV.   Thus, we see that the mass splitting varies from a
hundred MeV to a few GeV over most of parameter space.

Radiative corrections to this splitting can be calculated.  They arise
from corrections to the $\phi_+$ propagator due to loops involving the $Z$ and $\gamma$ and
corrections to the $\phi_S$ propagator due to $Z$ loops (the corrections due to the $W$
are identical in the limit that the masses of the $\phi_+$ and $\phi_S$ are the same, and
thus will be very small).  We have calculated these corrections and found that they are
seldom greater than  $250$ MeV.   However, loops due to the supersymmetric partners can also be
significant, and these will depend on gaugino/Higgsino masses and mixing parameters. 
Precise predictions can therefore not be made.   Nonetheless, this will not affect the
conclusion that the mass splitting tends to be $1-2$ GeV, but can be as low as a hundred
MeV and as high as a few GeV.

From an experimental point of view, the $\phi^+$ looks like a heavy,
charged lepton with an associated neutrino which is very slightly lighter.
Searches for heavy leptons with nearly degenerate neutrinos will be
sensitive to this particle.  (The fact that the $\phi$'s are scalars
will appear in the angular distribution in the production rate, as we
will see).   Recently, DELPHI\cite{delphi} has conducted a search for
charginos which are nearly mass-degenerate with the lightest
neutralino.  The signatures here will be very similar.  They show that
the standard chargino searches will cover mass splittings down to about
$3$ GeV, and that initial state radiation can extend this reach.  As we can see, this
will cover some, but not most, of parameter space.  Searches for ``stable" leptons will
also be sensitive if the mass splitting is less than $150$ MeV (above this, pion decays
dominate and the lifetime becomes substantially shorter).  In between, one must look
for the vertex, as with any intermediate range heavy lepton. 

The
fact that the model has so few parameters allows a precise
determination of the production cross-section, unlike the chargino case,
which depends on various mixing angles, scalar neutrino masses, etc.
The production cross-section for $\phi^+\phi^-$ at
an $e^+e^-$ collider is given by
\begin{equation}
\sigma={\pi\alpha^2\over 3s}\left(1-{4M_+^2\over
s}\right)^{3/2}\left[1+{A\over (1-{M_Z^2\over s})^2}\right]
\end{equation}
where $A\equiv \frac{\cos^22\theta_W}{4\sin^42\theta_W} \sim  0.15 $, and
we have not included the vector coupling of the electron to the $Z$,
which is proportional to ${1\over 4}-\sin^2\theta_W$.   For
$\sqrt{s}\sim 200$ GeV, this gives a cross-section of $0.6$ picobarns
times the phase space factor.  The angular distribution is the usual
$\sin^2\theta$ distribution for scalar particles.

\section{U(2) Model}\label{section3}
In this section we consider relaxing the assumption that a symmetry forbids
tree level FCNC. We do this by giving the model an explicit flavor symmetry. 
A very successful and elegant flavor symmetry that has been considered in the
literature is the (horizontal) U(2) flavor symmetry~\cite{u2papers}. 
In the U(2) model, which only involves the two Higgs doublets of the MSSM,
the matter fields of the first and second generations transform 
as the components
of a doublet, while the third generation transforms as a singlet, i.e. if we
denote the matter fields by $\psi$, 
then $\psi = \psi_a + \psi_3$, where $a = 1, 2$.
The two Higgs doublets transform as singlets. 
The minimal model contains three flavon fields which are responsible, through 
their vacuum expectation values (vevs), for the breaking of
the flavor symmetry. The flavons consist of a doublet $\phi$, a triplet $S$, 
and a (antisymmetric) singlet $A$. The breaking occurs in two steps
\begin{eqnarray}\label{breakup}
U(2) \stackrel{\epsilon}\rightarrow U(1) \stackrel{\epsilon'}
\rightarrow \,\, nothing \, ,
\end{eqnarray}
where $\epsilon$ is the vev of $\phi$ and $S$, while $\epsilon'$
is the vev of $A$. We will consider the unified version~\cite{u2papers} in 
which the model is embedded in a SU(5) GUT. In this case, the flavon
fields also transform under SU(5), and a new flavon $\Sigma$ is introduced.
The flavons and their transformations are
\begin{eqnarray} \label{flavonssu5}
\phi \sim ({\bf 1}, {\bf 2}) \,\, , S \sim ({\bf 75}, {\bf 3}) \,\, , \\
A \sim ({\bf 1}, {\bf 1}) \,\, , \Sigma \sim ({\bf 24}, {\bf 1}) \, ,
\end{eqnarray}
where the numbers in parenthesis correspond to the transformations 
under SU(5) and U(2) respectively. They acquire the following vevs:
\begin{eqnarray}\label{vevs}
\nonumber
{\langle S \rangle\over M_f} & \approx & \left( \begin{array}{cc} 0 & 0 \\ 0 & \epsilon 
\end{array} \right) \, , \,\,{\langle \phi \rangle\over M_f} \approx \left( \begin{array}{c} 
0 \\ \epsilon \end{array} \right) \, , \\ \nonumber \\
{\langle A \rangle\over M_f} & \approx & \,\, \epsilon^{ab} \,\, \epsilon' \,\,\,\,\,\, , 
\,\,\,\,\,\, {\langle \Sigma \rangle\over M_f} \approx \epsilon \, .
\end{eqnarray}
where $M_f\equiv \epsilon M_{GUT}$ is the flavor scale.
This model successfully reproduces the observed quark mass ratios and CKM angles, 
as well as the lepton mass ratios~\cite{u2papers}.
Let's now consider the possibility of having three Higgs generations and 
letting them transform non-trivially under the flavor symmetry~\cite{carone}.  
The matter fields are in ${\bf \bar{5}}$ $(\bar{F})$, and ${\bf 10}$'s
$(T)$ of SU(5). Then, their transformation
properties are $({\bf 10}, {\bf 2} \oplus {\bf 1})$ 
and $({\bf \bar{5}}, {\bf 2} \oplus {\bf 1})$ respectively, where 
again the first
term in the parenthesis corresponds to SU(5) and the second to U(2).
There are six Higgs doublets with components $H$ and $\bar{H}$ transforming
as  $({\bf 5}, {\bf 2} \oplus {\bf 1})$ and  $({\bf \bar{5}}, {\bf 2} 
\oplus {\bf 1})$ respectively.

The Yukawa part of the superpotential is
\begin{eqnarray}\label{WY}
W_Y & = & T_3 H_3 T_3 + T_3 \bar{H}_3 \bar{F}_3 \, \xi + 
\frac{1}{M_f} \left[ T_3 H_a \phi^a T_3 + T_3 \phi^a H_3 T_a
      + \left( T_3 \phi^a \bar{H}_3 \bar{F}_a + T_a \phi^a \bar{H}_3 \bar{F}_3
      \nonumber \right. \right. \\  
    & + & \left. \left. T_3 \phi^a \bar{H}_a \bar{F}_3
        + T_a (S^{ab}+A^{ab})\bar{H}_3 \bar{F}_b
        + T_a (S^{ab}+A^{ab}) \bar{H}_b \bar{F}_3
        + T_3( S^{ab}+A^{ab}) \bar{H}_a \bar{F}_b \right) \xi \right]
      \nonumber \\
    & + & \frac{1}{M_f^2} 
          \left[ T_a (S^{ab} \Sigma + A^{ab} \Sigma + \phi^a \phi^b) H_3 T_b 
         + T_3 (S^{ab} \Sigma + A^{ab} \Sigma + \phi^a \phi^b) H_a T_b 
          \nonumber \right. \\
    & + & \left. \left( T_3 \bar{H}_a \phi^a \phi^b \bar{F}_b
        + T_a \phi^a \phi^b \bar{H}_b \bar{F}_3
        + T_a (S^{ab} + A^{ab}) \phi^c \bar{H}_b \bar{F}_c \right) \xi \right] 
          \nonumber \\
    & + & \frac{1}{M_f^3} \left[ T_a \phi^a \phi^b \phi^c H_b T_c
        + T_a \phi^a \phi^b \phi^c \bar{H}_b \bar{F}_c \, \xi \right]\, ,
\end{eqnarray}
where $\xi \sim
m_b/m_t$.

The Yukawa coupling matrices can now be obtained from Eq.~(\ref{WY}) and
Eq.~(\ref{vevs}), their
textures are given by
\begin{eqnarray} \label{yu}
Y_U \approx \left( \begin{array}{ccc} 
0 & \epsilon \epsilon' & 0 \\
-\epsilon \epsilon' & \epsilon^2 & \epsilon \\
0 & \epsilon & 1 \end{array} \right)  H_3 
+ \left( \begin{array}{ccc} 
0 & \epsilon \epsilon' & \epsilon \epsilon' \\
-\epsilon \epsilon' & \epsilon^3 & \epsilon^2   \\
-\epsilon \epsilon' & \epsilon^2  & \epsilon
\end{array} \right) H_2
+ \left( \begin{array}{ccc} 
0 & 0 & 0 \\
0 & 0 & -\epsilon \epsilon'   \\
0 & \epsilon \epsilon'  & 0
\end{array} \right) H_1 \, ,
\end{eqnarray}
\begin{eqnarray} \label{yd}
Y_D \approx \left( \begin{array}{ccc} 
0 & \epsilon' & 0 \\
-\epsilon' & \epsilon & \epsilon \\
0 & \epsilon & 1 \end{array} \right) \xi  \bar{H}_3 
+ \left( \begin{array}{ccc} 
0 & \epsilon \epsilon' & \epsilon' \\
-\epsilon \epsilon' & \epsilon^2 & \epsilon \\
-\epsilon' & \epsilon & \epsilon
\end{array} \right) \xi \bar{H}_2
+ \left( \begin{array}{ccc} 
0 & 0 & 0 \\
0 & -\epsilon \epsilon' & -\epsilon' \\
0 & \epsilon' & 0
\end{array} \right) \xi \bar{H}_1 \, ,
\end{eqnarray}
where $O(1)$ coefficients have been omitted. 
Using the values $\epsilon \approx 0.02$ and $\epsilon' \approx 0.004$ obtained
from fitting the quark masses and mixing angles~\cite{u2papers}, one can 
calculate the contribution to FCNC.

A comprehensive analysis of flavor-changing processes in models with tree-level
FCNC was performed in Ref.~\cite{sheryuan} and Ref.~\cite{atwood}.   In
Ref.~\cite{chengsher}, it was argued that it is natural to take the flavor-changing
coupling $q_i q_j \phi$ to be the geometric mean of the
$q_i$ and
$q_j$ Yukawa couplings; this ansatz (referred to as ``Model III") has been used in a large
number of analyses of FCNC.  In Model III,  the strongest bound comes from
$K^0-\bar{K}^0$  mixing; the mass of the exchanged scalar has to be greater than
approximately a TeV. Much weaker bounds came from processes involving heavier quarks.  In many of these analyses, there has been a greater focus on FCNC between
the second and third generations; authors have generally assumed that the bound from
$K^0-\bar{K}^0$ mixing can be suppressed by some unspecified mechanism or a small amount
of fine tuning.   In the model of this Section the Yukawa couplings are different; the
$d
\, s
\, H_2$  coupling is $O(\epsilon \epsilon' \xi)$ which is $\sqrt{\epsilon'}$ times the
geometric mean of the down and strange Yukawa couplings. This significantly
weakens the bound; we find a bound on the $H_2$ mass of $17$ GeV from $K^0-\bar{K}^0$ mixing. In the case
of $B^0-\bar{B}^0$ mixing we obtain a bound of $100$ GeV (with $f_B \approx 200$ 
MeV). The bound coming from $D^0-\bar{D}^0$ mixing is of $\sim 120$
GeV. Bounds on processes involving leptons and b-quarks are much weaker.
Since $D^0-\bar{D}^0$ mixing is negligible in the standard model,
the first signature of this model could come from $D^0-\bar{D}^0$ mixing.

A similar analysis to the one described in the previous sections could
now be made. In this case however, the ``extra'' Higgs bosons do acquire
vevs and couple to fermions, as can be seen from Eqs.~(\ref{yu}$-$\ref{yd}).
We do not perform a detailed study of the phenomenology of the additional
Higgs scalars. 
\section{Conclusion}\label{section6}
We study the supersymmetric standard model with more than one generation of Higgs
doublets. We follow Ref.~\cite{griestsher} where a symmetry that forbids 
tree-level FCNC has been assumed. A result of this assumption is that the
additional Higgs bosons do not mix with the standard Higgs bosons, do not
couple to fermions, and that there is a mass 
degeneracy among the lightest neutral bosons. 
In this paper, it is noted
that the charged scalar is
heavier than the neutral scalar, for most of parameter space, by between a few hundred MeV and a few GeV, and
that standard searches for heavy leptons with nearly degenerate neutrinos may
be sensitive to these particles.   Lastly, we relax the assumption that tree-level FCNC
are eliminated, and discuss a flavor symmetry based on U(2) showing how the
level of FCNC processes can be related to quark masses, and how the FCNC coupling between
the first and second generations is automatically suppressed relative to the conventional
ansatz, significantly weakening bounds from  $K^0-\bar{K}^0$ mixing 

{\samepage
\begin{center}
{\bf Acknowledgments}
\end{center}
We thank Christopher D. Carone for useful conversations and for reading the
manuscript.   
We thank the National Science Foundation for support under 
Grants Nos.\ PHY-9800741 and PHY-9900657}
\end{document}